\documentclass[prd,eqsecnum,twocolumn,amsfonts,amssymb]{revtex4}

\usepackage{graphicx}

\usepackage{bm}

\newcommand{\beq}{\begin{equation}}
\newcommand{\eeq}{\end{equation}}
\newcommand{\beqs}{\begin{eqnarray}}
\newcommand{\eeqs}{\end{eqnarray}}

\begin{document}

\title{Comparison of Some Exact and Perturbative Results for a Supersymmetric 
SU($N_c$) Gauge Theory}

\author{Thomas A. Ryttov$^a$}
%\thanks{ryttov@physics.harvard.edu}

\author{Robert Shrock$^b$}
%\thanks{robert.shrock@stonybrook.edu}

\affiliation{(a) 
Department of Physics, Harvard University, Cambridge, MA  02138}

\affiliation{(b)
C.N. Yang Institute for Theoretical Physics, Stony Brook University,
Stony Brook, NY 11794}

\begin{abstract}

We consider vectorial, asymptotically free ${\cal N}=1$ supersymmetric
SU($N_c$) gauge theories with $N_f$ copies of massless chiral superfields in
various representations and study how perturbative predictions for the lower
boundary of the infrared conformal phase, as a function of $N_f$, compare with
exact results. We make use of two-loop and three-loop calculations of the beta
function and anomalous dimension of the quadratic chiral superfield operator
product for this purpose. The specific chiral superfield contents that we
consider are $N_f$ copies of (i) $F+\bar F$, (ii) $Adj$, (iii) $S_2+\bar S_2$,
and (iv) $A_2 + \bar A_2$, where $F$, $Adj$, $S_2$, and $A_2$ denote,
respectively, the fundamental, adjoint, and symmetric and antisymmetric rank-2
tensor representations.  We find that perturbative results slightly
overestimate the value of $N_{f,cr}$ relative to the respective exact results
for these representations, i.e., slightly underestimate the interval in $N_f$
for which the theory has infrared conformal behavior. Our results provide a 
measure of how closely perturbative calculations reproduce exact results
for these theories.

\end{abstract}


\maketitle

% ======================================================================

\section{Introduction} 

A longstanding question in gauge theories concerns how well the beta function,
calculated to same order in a perturbation expansion in the gauge coupling,
describes the properties of an asymptotically free gauge theory, in particular,
its evolution from high Euclidean momentum scales $\mu$ in the ultraviolet (UV)
to low $\mu$ in the infrared (IR).  Part of the difficulty in answering this
question stems from the fact that only the first two terms (i.e., the one-loop
and two-loop terms) in the beta function are scheme-independent.  Recall that
the beta function effectively resums the naive perturbation expansion in
defining a running gauge coupling $g(\mu)$.  There is special interest in the
case where the two-loop beta function has a zero away from the origin.  If this
zero occurs at a very small value of $\alpha(\mu) = g(\mu)^2/(4\pi)$, then one
expects that the theory does not confine or produce bilinear fermion
condensates and associated spontaneous chiral symmetry breaking (S$\chi$SB).
In contrast, if the theory has sufficiently few fermions, then this IR zero of
the beta function occurs at sufficiently large $\alpha$ that one expects the
theory to confine and spontaneously break chiral symmetry. For ordinary
non-supersymmetric SU($N_c$) gauge theories, based on the calculations of the
one-loop \cite{b1} and two-loop \cite{b2} terms in the beta function, there
have been a number of studies of this UV to IR evolution (an early work is
\cite{bz}).  A particularly interesting possibility is that the IR zero of the
beta function could occur at a value only slightly larger than the critical
value for S$\chi$SB, so that the theory would remain quasi-conformal for a
large interval in $\ln \mu$, with a large but slowly running (``walking'')
coupling and an associated large anomalous dimension $\gamma_m$ of the bilinear
fermion operator $\bar\psi\psi$ \cite{chipt,wtc}.  There has been intensive
recent work to study quasi-conformal behavior using lattice methods, and much
progress has been made \cite{lgt}.  These studies have considered not only
fermions in the fundamental representation of the gauge group (usually SU(2) or
SU(3)), but also fermions in higher-dimensional representations.  This has
reflected interest in quasi-conformal behavior in gauge theories
with fermions in higher-dimensional representations; for a review, see
\cite{sanrev}. 

In this paper we consider vectorial, asymptotically free ${\cal N}=1$
supersymmetric SU($N_c$) gauge theories (at zero temperature and chemical
potential) with content consisting of $N_f$ copies of massless chiral
superfields $\Phi_i$ and $\tilde \Phi_i$, $i=1,...,N_f$, which transform
according to representations $R$ and $\bar R$, respectively, where $R$ denotes
the representation of the gauge group.  We will present various results for
arbitrary $R$ and will analyze the following specific representations:
fundamental, adjoint, and rank-2 symmetric and antisymmetric tensor, denoted
$F$, $Adj$, $S_2$, and $A_2$, respectively. Using two- and three-loop
perturbative results, we study the evolution of the theory from the deep UV to
the IR and compare these perturbative results with exact results
\cite{seiberg94}-\cite{ryttov07}.  We investigate how the IR zero of the beta
function, calculated to a certain loop order, and the mass anomalous dimension,
$\gamma_m$, of the composite superfield operator product that contains the
component-field $\bar\psi\psi$, evaluated at this IR zero, calculated to the
same loop order, compare with exact results.  If the theory evolves from the UV
to a conformally invariant IR phase (a non-Abelian Coulomb phase), there is a
rigorous upper bound on $\gamma_m$, namely $\gamma_m \le 1$
\cite{mack75}-\cite{dobrev85}.  We apply this to a scheme-independent
calculation of $\gamma_m$ and to the two-loop and three-loop values of
$\gamma_m$ at the IR zeros of $\beta$, calculated to the same order, to
calculate perturbative estimates of the minimal value of $N_f$, denoted
$N_{f,cr}$, for which the IR behavior of the theory is conformal.  We carry out
this analysis in full detail for the case of $\Phi_i$, $\tilde \Phi_i$ in the
$F + \bar F$ representation and give briefer analyses for higher
representations. By comparing these perturbative predictions for $N_{f,cr}$
with exact results, we obtain a measure of the accuracy of the perturbative
analysis.  A related way to estimate $N_{f,cr}$ is from an approximate solution
of the Dyson-Schwinger equations for the relevant propagators.  The comparison
of the resultant predictions with exact results for the fundamental
representation was given in the important paper \cite{ans98}.  Our work
complements Ref. \cite{ans98}, since we analyze the perturbative $\gamma_m$
rather than approximate solutions to Dyson-Schwinger equations.

There are several motivations for this work.  First, it is always of
fundamental field-theoretic interest to compare how well a perturbative
calculation reproduces an exact result.  This is especially true in a quantum
field theory in view of the fact that a pertubative expansion is not a Taylor
series expansion, with finite radius of convergence, but instead, only
asymptotic, with zero radius of convergence \cite{az}.  This is to be
contrasted, for example, with high-temperature series expansions in statistical
mechanics and strong-bare-coupling expansions in lattice gauge theory, which
are true Taylor series with (at least) finite radii of convergence. Secondly,
given the great current interest in the nature of the UV to IR evolution of
asymptotically free gauge theories, as a function of their fermion content, it
is valuable to have a quantitative measure of the accuracy of the
(semi)perturbative approach of calculating the IR zero of the beta function and
evaluating the anomalous dimension evaluated at this zero of $\beta$.  We have
previously carried out such a study for non-supersymmetric SU($N_c$) gauge
theories with various fermion contents \cite{bvh} (see also \cite{ps}, whose
results are in agreement with those in \cite{bvh}).  An advantage of making
this comparison in a supersymmetric gauge theory is that, in contrast to the
non-supersymmetric case, one can compare the perturbative calculation with
exact results on the IR properties of the theory.

This paper is organized as follows.  In Sect. II we define some notation and
recall the relevant coefficients of the beta function. Sect. III is devoted to
a discussion of the perturbative expression for $\gamma_m$. In Sect. IV we
review some exact results on the IR phase structure of the theory. In Sect. V
we give a general discussion of perturbative estimates of $N_{f,cr}$.  The 
subsequent sections VI-XIII contain our results for the various representations
considered here.  Our conclusions are in Sect. IX. 

% ====================================================================

\section{Beta Function} 

\subsection{General} 

The beta function of the theory is denoted $\beta = dg/dt$, where $dt = d\ln
\mu$. In terms of the variable
\beq
a \equiv \frac{g^2}{16\pi^2} = \frac{\alpha}{4\pi} \ , 
\label{a}
\eeq
the beta function can be written equivalently as $\beta_\alpha \equiv
d\alpha/dt= g \beta/(2\pi)$, expressed as a series 
\beq
\frac{d\alpha}{dt} = -2\alpha \sum_{\ell=1}^\infty b_\ell \, a^\ell  
         = -2\alpha \sum_{\ell=1}^\infty \bar b_\ell \, \alpha^\ell \ , 
\label{beta}
\eeq
where $\ell$ denotes the number of loops involved in the calculation of
$b_\ell$ and $\bar b_\ell = b_\ell/(4\pi)^\ell$.  The first two
coefficients in the expansion (\ref{beta}), which are scheme-independent, are
\cite{jones75}
\beq
b_1 = 3C_A - 2T_f N_f
\label{b1}
\eeq
and \cite{machacek83}-\cite{jonesmez84}
\beq
b_2=6C_A^2-4(C_A+2C_f)T_fN_f \ . 
\label{b2}
\eeq
A commonly used regularization scheme for supersymmetric theories is
dimensional reduction with minimal subtraction, denoted $\overline{DR}$
\cite{dred,ggrs} (a recent discussion is \cite{stockinger05}). In this
regularization scheme, the coefficient of the three-loop term in the beta
function is \cite{jjn96a}-\cite{hms09}
\beqs
b_3 & = & 21C_A^3 + 4(-5C_A^2-13C_AC_f+4C_f^2)T_fN_f \cr\cr
    & + & 4(C_A+6C_f)T_f^2N_f^2 \ . 
\label{b3}
\eeqs

Although the beta function coefficients $b_\ell$ with $\ell \ge 3$ are
scheme-dependent, the use of three- and four-loop coefficients in the
comparison of the QCD beta function with experimental data has shown the value
in incorporating these higher-loop contributions \cite{bethke}.  The purpose in
our present work is different from that for QCD; there, one wanted
to obtain the most precise comparison possible with data, in order to extract
the value of $\alpha_s(\mu)$.  Here, we would like to compare the predictions
of a particular scheme, namely DR, with the exact results concerning the
infrared behavior of the theory for various ranges of $N_f$, in order to obtain
a measure of the accuracy and reliability of the perturbative calculations as a
guide to the infrared phase structure of the theory. 

Before proceeding, it is appropriate to include several cautionary remarks.
First, as is well-known, expansions such as (\ref{beta}) and (\ref{gamma}) in
powers of $\alpha$ are not Taylor series, but instead, only asymptotic series,
with zero radius of convergence.  However, a wealth of experience in particle
physics has shown that if the effective expansion parameter (here,
$(\alpha/\pi)$ times various group invariants) is not too large, then the first
few terms can provide useful information about the physics.  Second, the
expansions for $\beta$ and $\gamma_m$ in Eqs.  (\ref{beta}) and (\ref{gamma})
are perturbative and do not incorporate nonperturbative properties of the
physics, such as instantons.  Instanton effects are absent to any order of a
perturbative expansion in $\alpha$, but play an important role in a non-Abelian
Yang Mills gauge theory.  Terms arising from instanton effects
characteristically involve essential zeros of the form ${\rm exp}(-\kappa
\pi/\alpha)$, where $\kappa > 0$ is a numerical constant.  Indeed,
instanton effects play an important role in the derivation of exact results on
supersymmetric gauge theories \cite{seiberg94},\cite{nsvz}. 

The requirement that the theory be asymptotically free means that $\beta < 0$,
which, with the overall minus sign in Eq. (\ref{beta}), is true if and only if
$b_1 > 0$.  Note that, {\it a priori}, the condition $\beta > 0$ could be
satisfied with $b_1=0$ if $b_2 > 0$, but this is actually not possible, because
the value of $N_f$ that renders $b_1=0$, namely \cite{nfintegral}
\beq
N_{f,b1z} = \frac{3C_A}{2T_f} \equiv N_{f,max} \ , 
\label{nfb1z}
\eeq
yields a negative value of $b_2$ for any representation $R$, viz., 
$-12C_A C_f$. Hence, the requirement of asymptotic freedom implies
\beq
N_f < N_{f,max} \ . 
\label{nfmax}
\eeq
The number $N_{f,max}$ depends on the representation $R$, and, where
necessary for clarity, we shall indicate this by writing $N_{f,max,R}$.  We
shall assume $N_f$ satisfies this upper bound in our present study.  Although
one could generalize the analysis to non-asymptotically free theories, the
coupling $\alpha(\mu) \to 0$ as the energy scale $\mu \to 0$ in such theories,
so the infrared behavior would be that of a free theory.

\subsection{Zero of the Two-Loop Beta Function}

Since only the two-loop beta function is scheme-independent, at
the perturbative level, if it does not have an IR zero, then, even if such a
zero were present at the level of three or more loops, it could not be reliably
considered to be physical.  Here we have an even more stringent criterion,
based on the exact results of Ref. \cite{seiberg94}, which specify, as a
function of $N_c$ and the chiral fermion content, whether the theory evolves
from the UV to a conformally invariant IR phase (a non-Abelian Coulomb phase) .
These results are equivalent to having an exact beta function and knowing
whether it has an exact IR fixed point of the renormalization group. If the
exact analysis does not have an IR zero but the perturbative 2-loop beta
function does have an IR zero, then even though the latter is
scheme-independent, one would still have to reject its prediction, since it
differs from the exact result.  In Ref. \cite{btd}, a nonperturbative IR
zero of the beta function of a non-Abelian gauge theory has also been
discussed.

For zero and sufficiently small $N_f$, the coefficients $b_2$ and $b_3$ are 
both positive.  As $N_f$ increases, these coefficients both decrease.  The
coefficient $b_2$ passes through zero and reverses sign from positive to
negative at the value 
\beq
N_{f,b2z} = \frac{3C_A^2}{2T_f(C_A+2C_f)} \ . 
\label{nfb2z}
\eeq
This value of $N_f$ is less than $N_{f,b1z}=N_{f,max}$, as is clear from the 
fact that 
\beq
N_{f,b2z} = \frac{N_{f,b1z}}{1+\frac{2C_f}{C_A}} < N_{f,b1z} \ .
\label{Nfb2zltNfb1z}
\eeq

The two-loop ($2\ell$) beta function has a zero away from the origin at
$a_{IR,2\ell}=-b_1/b_2$, i.e., 
\beqs
\alpha_{IR,2\ell} & = & -\frac{4\pi b_1}{b_2} \cr\cr
       & = & \frac{2\pi(3C_A-2T_fN_f)}{2(C_A+2C_f)T_fN_f-3C_A^2} \ . 
\label{alfir2loop}
\eeqs
Clearly, for $N_f$ only slightly larger than $N_{f,b2z}$, $\alpha_{IR,2\ell}$
is too large for this perturbative result to be trustworthy; a necessary
condition for it to be reliable is that $N_f$ is sufficiently far above
$N_{f,b2z}$ that $\alpha_{IR,2\ell}$ is not too large.  For our analysis below,
it will be important whether the formal divergence in $\alpha_{IR,2\ell}$ at
$b_2=0$, i.e., $N_f=N_{f,b2z}$, occurs above or below the lower boundary of the
IR conformal phase, which is given by $N_{f,cr}$ in Eq. (\ref{nfcr}).  The
difference is
\beq
N_{f,b2z}-N_{f,cr} = \frac{3C_A(C_A-2C_f)}{4T_f(C_A+2C_f)} \ . 
\label{nfb2zminusnfcr}
\eeq
We find that this can be positive or negative.  For example, for the
fundamental representation, 
\beq
N_{f,b2z}-N_{f,cr} = \frac{3N_c}{2(2N_c^2-1)} > 0  \quad {\rm for \ fund. \
  rep.} \ , 
\label{nfb2zminusnfcr_fund}
\eeq
so that $b_2=0$ and $\alpha_{IR,2\ell}$ diverges within the IR conformal 
phase.  In contrast, for the adjoint representation, 
\beq
N_{f,b2z}-N_{f,cr} = - \frac{1}{4} \quad {\rm for \ Adj. \ rep.} \ , 
\label{nfb2zminusnfcr_adj}
\eeq
so that in this case, $b_2$ is nonzero (and negative) all throughout the IR
conformal phase.  For the symmetric and antisymmetric rank-2 tensor
representations, we find
\beqs
N_{f,b2z}-N_{f,cr} & = & - \frac{3N_c(N_c^2 \pm 2N_c-4)}{2(N_c \pm 2)
(3N_c^2 \pm 2N_c-4)} \cr\cr
& & {\rm for} \ S_2, \ A_2 \ {\rm rep.} \ , 
\label{nfb2zminusnfcr_t2}
\eeqs
where the upper and lower signs apply for the $S_2$ and $A_2$ representations,
respectively.  For $S_2$, the numerator factor $N_c^2+2N_c-4$ vanishes at the
unphysical, negative value $N_c=-(1+\sqrt{5} \ )$ and at $N_c=-1+\sqrt{5}
\simeq 1.236$, which is less than the minimal non-Abelian value, $N_c=2$.
Hence, for the $S_2$ representation, $N_{f,b2z} < N_{f,cr}$ for all non-Abelian
$N_c$ and $b_2$ has fixed (negative) sign throughout the IR conformal phase.
For the $A_2$ representation, $N_c$ is restricted to the nontrivial range $N_c
\ge 3$.  In this $A_2$ case, the numerator factor $N_c^2-2N_c-4$ vanishes at
$N_c=1+\sqrt{5} \simeq 3.236$ (as well as at the negative, unphysical value
$N_c=1-\sqrt{5}$), so that $N_{f,b2z} > N_{f,cr}$ for the real interval $3 \le
N_c < 1+\sqrt{5}$, while $N_{f,b2z} < N_{f,cr}$ for $N_c > 1+\sqrt{5}$, i.e.,
the integer values $N_c \ge 4$. Note that the $A_2$ representation with $N_c=3$
is equivalent to the conjugate fundamental representation.  Hence, for all
representations $R$ for which $A_2$ is distinct from the fundamental, $b_2$ has
fixed (negative) sign throughout the IR conformal phase.

Given that $N_f < N_{f,max}$ to maintain the asymptotic freedom of the theory,
this $\alpha_{IR,2\ell}$ is positive and hence physical if and only if $N_f$
lies in the range $N_{f,b2z} < N_f < N_{f,b1z}$, i.e.,
\beq
\frac{3C_A^2}{2T_f(C_A+2C_f)} < N_f < \frac{3C_A}{2T_f} \ . 
\label{nfinterval}
\eeq
We will thus focus on this interval for $N_f$.  The zero of the two-loop beta
function at $\alpha=\alpha_{IR,2\ell}$ is either an approximate or exact
infrared (IR) fixed point (IRFP) of the renormalization group.  If the gauge
interaction spontaneously breaks the global chiral symmetry of the theory via
the formation of a bilinear matter (chiral) superfield condensate, then this IR
zero is only approximate, since in this case the matter superfield picks up a
dynamically generated mass $\Sigma$ and as the scale $\mu$ decreases below
$\Sigma$, one integrates out the matter superfields in defining the effective
low-energy field theory.  Consequently, as the theory evolves further into the
infrared, the beta function becomes that of the pure supersymmetric gauge
theory without these matter superfields, and hence $\alpha(\mu)$ evolves away
from the approximate IR fixed point.

\subsection{Zeros of the Three-Loop Beta Function}

To three-loop order, the beta function formally has two zeros away from the
origin, given by the equation $b_1+b_2a+b_3a^2=0$, where $a$ was given in 
Eq. (\ref{a}). The solutions, in terms of $\alpha= 4\pi a$, are
\beq
\alpha = \frac{2\pi}{b_3}\Big [ -b_2 \pm \sqrt{b_2^2-4b_1b_3} \ \Big ] \ . 
\label{alphazerogen}
\eeq
Only the physical, smaller one of these two solutions will be relevant for our
analysis, and we label it as $\alpha_{IR,3\ell}$.  As discussed above, the
requirement that the two-loop beta function has an IR zero means that $N_f$ is
in the interval (\ref{nfinterval}) where $b_1 > 0$ and $b_2 < 0$.

% ======================================================================

\section{Anomalous Dimension $\gamma_m$ }

The anomalous dimension $\gamma_m$ describes the scaling properties of the
quadratic superfield operator product $\Phi_i \tilde\Phi_i$ containing the
bilinear product $\psi^T C \tilde \psi$, or equivalently, $\bar\psi\psi$, of
component fermion fields.  If one has an input mass $m$ for $\psi$, then, with
our definition, $\gamma_m = -d\ln m/dt$, where $t=\ln \mu$.  Since we are
studying the evolution from the UV to the IR conformal phase, we do not put in
such a bare mass $m$ here, since, if we did, then as $\mu$ decreases below $m$,
these fields would be integrated out as the theory evolved deeper into the
infrared and the IR behavior would be that of a supersymmetric SU($N_c$) theory
with just gluons and gluinos. For notational simplicity we will often suppress
the subscript $m$. This anomalous dimension can be expressed as a series in $a$
or equivalently, $\alpha$:
\beq
\gamma_m = \sum_{\ell=1}^\infty c_\ell \, a^\ell 
   = \sum_{\ell=1}^\infty \bar c_\ell \, \alpha^\ell \ , 
\label{gamma}
\eeq
where $\bar c_\ell = c_\ell/(4\pi)^\ell$ is the $\ell$-loop series coefficient.
We denote $\gamma_{n\ell}$ as the $n$-loop value of $\gamma_m$, i.e., 
$\gamma_{n\ell} = \sum_{n=1}^\ell c_n \, a^n$.

The coefficients $c_\ell$ have been calculated to three-loop order. The
one-loop coefficient $c_1$ is scheme-independent: 
\beq
c_1 = 4C_f \ , 
\label{c1}
\eeq
The higher-loop coefficients $c_\ell$ with $\ell \ge 2$ are scheme-dependent.
In the $\overline{DR}$ scheme, $c_2$ and $c_3$ are \cite{jjn96b,hms09} 
\beq
c_2 = 4C_f(-2C_f+3C_A-2T_fN_f) \ , 
\label{c2}
\eeq
and
\begin{widetext}
 \beq
c_3 = 8C_f\Bigg [ 4C_f^2+3C_A(C_A-C_f) + T_fN_f\Big [ (-8+12\zeta(3))C_f
+(1-12\zeta(3))C_A \Big ] -2T_f^2N_f^2 \Bigg ] \ , 
\label{c3}
\eeq
\end{widetext}
where $\zeta(s)$ is the Riemann zeta function, with $\zeta(3) = 
1.20205690..$  As $N_f$ approaches
$N_{f,max}$ from below, $b_1 \to 0$ with nonzero $b_2$ and hence
$\alpha_{IR} \to 0$; since the perturbative calculation expresses $\gamma_m$ in
a power series in $\alpha$, it follows that $\gamma_m \to 0$ as $N_f \to
N_{f,max}$.

The $n$-loop value of $\gamma_m$ at the IR zero of $\beta$, calculated to the
same loop order (IR fixed point of the renormalization group), is obtained by
setting $\alpha = \alpha_{IR,n\ell}$ in $\gamma_{n\ell}(\alpha)$ and is 
denoted 
\beq
\gamma_{IR,n\ell} \equiv \gamma_{n\ell}(\alpha_{IR,n\ell}) \ .
\label{gammirnell}
\eeq
where the dependence on the chiral superfield representation $R$ is implicit. 
Thus, at the two-loop level, 
\beqs
\gamma_{IR,2\ell} & = & a(c_1+c_2a)|_{a=a_{IR,2\ell}} \cr\cr
                  & = & \frac{b_1(-c_1 b_2 + c_2 b_1)}{b_2^2} \ .
\label{gammair2loop}
\eeqs
Explicitly, 
\begin{widetext}
\beq
\gamma_{IR,2\ell} = \frac{C_f(3C_A-2T_fN_f)(2T_fN_f-C_A)(2T_fN_f-3C_A+6C_f)}
{[2(C_A+2C_f)T_fN_f-3C_A^2]^2} \ .
\label{gammair2loopexplicit}
\eeq
\end{widetext}
Thus, $\gamma_{IR,2\ell}$ has, formally, three zeros and one pole.  One of the
zeros occurs at $N_f=N_{f,max}$, as given in Eq. (\ref{nfb1z}).  The second
zero occurs at 
\beq
N_f = \frac{C_A}{2T_f} = \frac{N_{f,max}}{3} = \frac{2N_{f,cr}}{3} \ . 
\label{nfgamm2loopz1}
\eeq
Because this lies below the exact $N_{f,cr}$, it is not directly relevant for 
our current analysis.  Furthermore, for the representations of interest here,
it also lies below $N_{f,b2z}$, and hence is not present where the theory has a
two-loop zero in $\beta$.  The third formal zero in $\gamma_{IR,2\ell}$ occurs
at 
\beq
N_f = \frac{3(C_A-2C_f)}{2T_f} = N_{f,max} - \frac{3C_f}{T_f} \ .
\label{nfgamma2loopz2}
\eeq
This third zero occurs for $N_f$ less than $N_{f,cr}$ (and, for some
representations, at negative $N_f$), and hence also will not be relevant for
our analysis.  The pole in $\gamma_{IR,2\ell}$ occurs at $N_{f,b2z}$, and
is a consequence of the pole in $\alpha_{IR,2\ell}$ where $b_2=0$.  Clearly,
the two-loop calculation of $\gamma_m$ ceases to be reliable for $N_f$ values
greater than $N_{f,b2z}$, so this pole is obviously an unphysical artifact. 
Thus, over the range of interest here, $\gamma_{IR,2\ell}$ increases
monotonically above zero as $N_f$ decreases below $N_{f,max}$.

In the procedure described above, one evaluates the $n$-loop expression for
$\gamma_m$ at the IR zero of $\beta$, calculated to the same $n$-loop order.
For this procedure, one necessarily uses the two-loop $\beta$ function, since
an IR zero only appears at this loop level, and also the two-loop or
higher-loop expressions for $\gamma_m$. Since the coefficients $c_\ell$ for
$\ell \ge 2$ are scheme-dependent, this process necessarily entails
scheme-dependence.  We thus also present an alternate perturbative estimate for
$\gamma_m$, which has the advantage of preserving scheme-independence but the
disadvantage of mixing different orders of perturbation theory. For this
alternative estimate, we use only scheme-independent (SI) inputs, and hence
evaluate the one-loop expression for $\gamma_m$ at the two-loop IR zero of
$\beta$, obtaining
\beqs
\gamma_{IR,SI} & = &  c_1 a_{IR,2\ell} = \bar c_1 \alpha_{IR,2\ell} = 
-\frac{c_1 b_1}{b_2} \cr\cr
            & = & \frac{2C_f(3C_A-2T_fN_f)}{2(C_A+2C_f)T_fN_f-3C_A^2} \ .
\label{gammasi}
\eeqs
%

% =======================================================================

\section{Review of Some Exact Results}

Since we will compare our perturbative results with certain exact results, a
brief review of these is appropriate. For a vectorlike SU($N_c$) gauge theory
with ${\cal N}=1$ supersymmetry and $N_f$ copies of massless chiral superfields
$\Phi_i$ and $\tilde \Phi_i$ in the fundamental and conjugate fundamental
representation, respectively, exact results on the phase structure and
corresponding properties of the theory in the infrared were derived by Seiberg
\cite{seiberg94}. These results were subsequently generalized to theories with
gauge groups SO($N_c$) and Sp($N_c$) in \cite{intriligator95a} and
\cite{intriligator95b} (reviews include \cite{susyreviews}).  A further
generalization to arbitrary representations was given in \cite{ryttov07}.  In
our present work we will focus on the comparison of perturbative estimates and
exact results concerning the minimal value of $N_f$ (for a given chiral
superfield content), denoted $N_{f,cr}$ such that, for $N_f > N_{f,cr}$, the
theory evolves from the UV to the IR in a chirally symmetric manner, so that
the IR theory is a conformal, non-Abelian Coulomb phase. This value,
$N_{f,cr}$, is often called the lower end of the conformal phase or conformal
``window'' (with the upper end, $N_{f,max}$, determined by the requirement of
asymptotic freedom).  We shall carry out this comparison at the two- and
three-loop level.

We recall how, for a given $R$, the conformal region in $N_f$ is determined.
A crucial tool in determining $N_{f,cr}$, the lower end, as a function of
$N_f$, of the IR conformal phase, is the existence of an exact relation between
the beta function and the mass anomalous dimension, $\gamma_m$.  This relation
is embodied in the following form for the beta function of the
theory with a vectorlike massless chiral superfield content consisting of $N_f$
copies of the representations $R+\bar R$ of the gauge group \cite{nsvz,ansatz}:
\beq
\beta_\alpha = \frac{d\alpha}{dt} = - \frac{\alpha^2}{2\pi} \bigg [
\frac{b_1-2T_fN_f\gamma_m(\alpha)}
     {1-\frac{C_A \, \alpha}{2\pi} } \bigg ] \ . 
\label{beta_nsvz}
\eeq
The IR zero of $\beta_\alpha$ is determined by the condition 
\beq
\gamma_m = \frac{3C_A-2T_fN_f}{2T_fN_f} = \frac{N_{f,max}}{N_f} - 1 \ . 
\eeq
Let us assume that the theory flows to an exact IR fixed point, i.e., that
$N_f$ is in the sub-interval of (\ref{nfinterval}) in which, as the theory
evolves down from the UV to the IR, no spontaneous chiral symmetry breaking
takes place. Given that the theory has evolved down to an (exact) infrared
fixed point,the resultant theory at this IRFP is conformally invariant.  

It is a special property of a conformally invariant field theory (whether
supersymmetric or not) that the full dimension of a spinless operator (other
than the identity) must be larger than unity in order that the theory not
contain any negative-norm states, which would violate unitarity
\cite{mack75,flato83,dobrev85}. Specifically, for the dimension $D_m$ of the
bilinear operator $\Phi_i \tilde \Phi_i$ (with no sum on $i$) for any
$i=1,...,N_f$ in the present theory, this is the inequality
\beq
D_m \ge 1 \ . 
\label{dmbound}
\eeq
In terms of its component scalar and fermion fields $\phi_i$ and $\psi_i$, the
chiral superfield is expressed as $\Phi_i = \phi_i + \sqrt{2} \, \theta \psi_i
+ \theta \theta F_i$, where $\theta$ is a Grassmann variable and $F_i$ is an
auxiliary field.  Thus (for any $i$), the term $\Phi_i \tilde \Phi_i$ yields,
as the (holomorphic) term bilinear in component fermion fields, $\theta \theta
\psi_i \tilde \psi_i$. Taking into account that the dimension of $\theta$ is
$-1/2$, the free-field dimension of $\psi_i \tilde \psi_i$ (for any $i$) is 3,
and using our definition of $\gamma_m$, it follows that $D_m = 2-\gamma_m$, so
that the bound (\ref{dmbound}) is equivalent to the following upper bound on
$\gamma_m$:
\beq
\gamma_m \le 1 \ . 
\label{gammabound}
\eeq
This may be contrasted with the situation in a non-supersymmetric SU($N_c$)
theory.  There, the bound that the full operator dimension of
$\bar\psi_i\psi_i$ be larger than 1 implies that $\gamma_m \le 2$, as we noted
in Eq. (4.2) of \cite{bvh} (equivalent to the bound from Eq. (4.1) of
\cite{bvh}).  The more stringent upper bound (\ref{gammabound}) on $\gamma_m$
in the supersymmetric theory is due to the fact that the fermion field $\psi_i$
is part of a chiral superfield and the holomorphic fermion bilinear resulting
from the quadratic $\Phi_i \tilde \Phi_i$ product carries with it a $\theta
\theta$ factor.

We next assume that, in the relevant range of $N_f$ where the theory evolves
from the UV to an IR-conformal phase, $\gamma_m$ evaluated at the IR fixed
point, $\alpha=\alpha_{IR}$, increases monotonically as $N_f$ decreases below
$N_{f,max}$. This assumption is satisfied by $\gamma_m$ as calculated in a
scheme-independent manner, as will be discussed further below. The inequality
(\ref{gammabound}) then implies that $N_{f,cr}$, the value of $N_f$ below
which the theory cannot be conformally invariant, is bounded below as $N_{f,cr}
\ge 3C_A/(4T_f)$. The application of duality relations provides strong evidence
that this inequality is saturated \cite{seiberg94,ryttov07} and hence that
\beq
N_{f,cr} = \frac{3C_A}{4T_f} \ . 
\label{nfcr}
\eeq
We refer to this as an exact result, although, as we have indicated, there
are some nonrigorous steps in its derivation. Note that 
\beq
N_{f,cr} = \frac{N_{f,max}}{2} \ . 
\label{nfcrnfmaxrel}
\eeq
Thus, the theory evolves from the UV to an IR fixed point in the conformal
phase if and only if $N_{f,cr}$ lies in the interval
\beq
\frac{3C_A}{4T_f} < N_f < \frac{3C_A}{2T_f} \ . 
\label{nfconformal}
\eeq
(The marginal value $N_f = 3C_A/(4T_f)$ itself is not in this conformal phase
\cite{seiberg94,intriligator95a}.) For both Eqs. (\ref{nfcr}) and
(\ref{nfconformal}), it is understood that, physically, $N_f$ must be an
integer \cite{nfintegral}.  Thus, the actual values of $N_f$ in the conformal
phase are understood to be the integers that satisfy the inequality
(\ref{nfconformal}).

% ======================================================================

\section{Perturbative Estimates of $N_{f,cr}$ for general $R$}

As discussed above, although a perturbative calculation is not exact, one gains
valuable information by comparing it with exact results.  We carry out this
comparison here for a general representation $R$, using perturbative estimates
for $N_{f,cr}$, the lower boundary of the IR conformal phase.  For this
purpose, we utilize $\gamma_{IR,SI}$ and $\gamma_{IR,2\ell}$.  With a monotonic
increase in $\gamma_m$ as $N_f$ decreases below $N_{f,max}$, we can then
calculate a perturbative estimate for $N_{f,cr}$ by assuming that $\gamma_m$
saturates the inequality (\ref{gammabound}) as $N_f$ decreases through
$N_{f,cr}$. (Here, again, we are implicitly analytically continuing $N_f$ from
physical integer values to real numbers.)  Setting the perturbative $\gamma_m =
1$ and solving for the value of $N_f$ at which this happens yields the
corresponding perturbative estimate of $N_{f,cr}$.  Since a perturbatively
calculated expression for $\gamma_m$ is not, in general, equal to the exact
$\gamma_m$, one does not expect these estimates to agree precisely with the
exactly known values for $N_{f,cr}$ for the various representations. However,
this comparison gives quantitative insight as to the accuracy of the
perturbative calculations.

\subsection{Estimate Using $\gamma_{IR,SI}$} 

The scheme-independent perturbative result for $\gamma_m$, $\gamma_{IR,SI}$,
increases monotonically as $N_f$ decreases from its maximal value
(\ref{nfb1z}) and reaches the rigorous upper bound as $N_f$ decreases through
the value 
\beq
N_{f,cr,SI} = \frac{3C_A(C_A+2C_f)}{2T_f(C_A+4C_f)} \ . 
\label{nfcrsi}
\eeq
This is larger than the exact value of $N_{f,cr}$, as is evident from the
difference 
\beq
N_{f,cr,SI} - N_{f,cr} = \frac{3C_A^2}{4 T_f(C_A+4C_f)} > 0 
\label{nfcrsiminusnfcr}
\eeq
or the ratio
\beq
\frac{N_{f,cr,SI}}{N_{f,cr}} = 2 \bigg ( \frac{C_A+2C_f}{C_A+4C_f} \bigg ) > 1
\ . 
\label{nfcrsiratio}
\eeq
This difference between the scheme-independent perturbative estimate of the
lower boundary of the conformal phase, $N_{f,cr,SI}$, and the exact lower
boundary, $N_{f,cr}$, provides one quantitative measure of the accuracy of
perturbation theory.  Our conclusion from this comparison is that perturbation
theory slightly overestimates the value of this lower boundary and hence
underestimates the size of the conformal phase as a function of $N_f$.  Related
to this, as $N_f$ decreases below $N_{f,cr,SI}$ toward the exact lower boundary
of the IR conformal phase at $N_{f,cr}$, $\gamma_{IR,SI}$ continues to
increase.  In this regime, its behavior is unphysical since it violates the
rigorous bound (\ref{gammabound}).  This happens for both representations where
$N_{f,b2z} > N_{f,cr}$ and representations where $N_{f,b2z} < N_{f,cr}$.
Formally,
\beq
\gamma_{IR,SI} = \frac{2C_f}{2C_f-C_A} \quad {\rm at} \ \ N_f = N_{f,cr} \ .
\label{gammasi_nfcr}
\eeq

\subsection{Estimate Using $\gamma_{IR,2\ell}$} 

Setting the two-loop result for the anomalous dimension, $\gamma_{IR,2\ell}$,
equal to the rigorous upper bound, unity, we derive the corresponding two-loop
perturbative prediction for $N_{f,cr}$.  The equation $\gamma_{IR,2\ell}=1$ is
a cubic equation in $N_f$, which yields the resultant estimate for 
$N_{f,cr}$, together with two other roots that are not of direct
relevance.  Formally, 
\beq
\gamma_{IR,2\ell} = \frac{C_f(4C_f-C_A)}{2(2C_f-C_A)^2} \quad {\rm at} \ \ 
N_f = N_{f,cr} \ , 
\label{gamma2loop_nfcr}
\eeq
but as with $\gamma_{IR,SI}$, this is only formal, since this perturbative
result generically violates the upper bound (\ref{gammabound}).  We comment
below on the situation at the three-loop level.  We proceed to present results
for the various representations of interest here.

\section{Chiral Superfields in the Fundamental Representation}

\subsection{IR Zeros of the Beta Function}

\subsubsection{Two-Loop Analysis}

In this section we consider the case where the theory has $N_f$ copies of
massless chiral superfields $\Phi_i$ and $\tilde \Phi_i$, $i=1,...,N_f$, 
transforming according to
\beq
\Phi_i: \ F; \quad \tilde \Phi_i: \ \bar F \ , \quad i=1,...,N_f \ , 
\label{fundrep}
\eeq
i.e., the fundamental plus conjugate fundamental,
representation of the gauge group.  The requirement of asymptotic freedom 
implies 
\beq
N_f < 3N_c \ . 
\label{nfmax_fund}
\eeq
For this case, the exact result (\ref{nfcr}) on the value of $N_f$ at the 
lower boundary of the conformal phase in the infrared is
\beq
N_{f,cr} = \frac{3N_c}{2} \ , 
\label{nfcr_fund}
\eeq
where it is understood that this is only formal if $N_c$ is odd, since
$N_{f,cr}$ must be an integer. Thus, the IR conformal phase is given, from
Eq. (\ref{nfconformal}), as
\beq
\frac{3N_c}{2} < N_f < 3N_c \ . 
\label{nfconformal_fund}
\eeq
Physically, $N_f$ must be a (non-negative) integer, so the
actual physical values of $N_f$ in the IR conformal phase for $2 \le N_c \le 
5$ are $N_c=2: \quad N_f=4, \ 5$; $N_c=3: \quad  N_f=5, \ 6, \ 7, \ 8$; 
$N_c=4: \quad N_f=7, \ 8, \ 9, \ 10, \ 11$;, and 
$N_c=5: \quad N_f=8, \ 9, \ 10, \ 11, \ 12, \ 13, \ 14$.

Evaluating Eq. (\ref{nfb2z}), we find that $b_2$ reverses sign from positive to
negative as $N_f$ increases through the value \cite{nfintegral}
\beq
N_{f,b2z} = \frac{3N_c}{2-N_c^{-2}} \ . 
\label{nfb2z_fund}
\eeq
The interval of
$N_f$ values in Eq. (\ref{nfinterval}) where the two-loop beta function has an
IR zero is therefore \cite{nfintegral}
\beq
\frac{3N_c}{2-N_c^{-2}} < N_f < 3N_c \ . 
\label{nfinterval_fund}
\eeq
Numerical values of $N_{f,cr}$, $N_{f,b2z}$, $N_{f,b3z}$, and
$N_{f,b1z}=N_{f,max}$ are listed in Table \ref{nfbz_fund} for the illustrative
values $2 \le N_c \le 5$ \ \cite{nfintegral}.  As discussed before in
connection with Eq. (\ref{nfb2zminusnfcr_fund}), the value of $N_{f,cr}$ in
Eq. (\ref{nfb2z_fund}) is greater (for all $N_c$) than the exactly known lower
boundary of the conformal phase in Eq. (\ref{nfcr_fund}).  As $N_c \to \infty$,
$N_{f,b2z}$ asymptotically approaches $(3/2)N_c$ from above.

\begin{table}
\caption{\footnotesize{Values of $N_{f,b1z}=N_{f,max}$, $N_{f,b2z}$, and
$N_{f,b3z}$ for the supersymmetric SU($N_c$) theory with
$N_f$ chiral superfields $\Phi_i$, $\tilde\Phi_i$ in the $F$ and $\bar F$ 
representations, respectively.  We also list the exact value of
$N_{f,cr}$.  These results are given for the illustrtive values 
$2 \le N_c \le 5$.}}
\begin{center}
\begin{tabular}{|c|c|c|c|c|} \hline\hline
$N_c$ & $N_{f,cr}$ & $N_{f,b2z}$ & $N_{f,b3z}$ & $N_{f,b1z}$  \\ \hline
 2    & 3          &    3.43   &     3.09        &    6    \\
 3    & 4.5        &    4.76   &     4.27        &    9    \\
 4    & 6          &    6.19   &     5.55        &   12    \\
 5    & 7.5        &    7.65   &     6.85        &   15    \\
\hline\hline
\end{tabular}
\end{center}
\label{nfbz_fund}
\end{table}

Since $N_{f,b2z} > N_{f,cr}$, it follows that the two-loop beta function only
has a (perturbative) infrared zero for $N_f$ values in the interval where the
theory is conformally invariant. This is different from the non-supersymmetric
SU($N_c$) gauge theory, in which the two-loop beta function may have an IR zero
for values of $N_f$ less than the estimate, from the Dyson-Schwinger equation,
of $N_{f,cr}$, i.e., in the phase where the theory has spontaneous chiral
symmetry breaking.  (Because of this S$\chi$SB, this IR zero is only
approximate.) 

For our supersymmetric theory with the $F+\bar F$ chiral superfield content of
Eq. (\ref{fundrep}), the general formula (\ref{alfir2loop}) for the IR zero of
the two-loop beta function reduces to
\beq
\alpha_{IR,2\ell} = \frac{2\pi(3N_c-N_f)}{(2N_c-N_c^{-1})N_f-3N_c^2}
\label{alfir2loop_fund}
\eeq
This decreases monotonically from arbitrarily large values (where, of course,
the perturbative beta function does not apply reliably) to zero as $N_f$
increases throughout the interval (\ref{nfinterval_fund}). Numerical values of 
$\alpha_{IR,2\ell}$ are listed in Table \ref{betazero_fund} for the
illustrative cases $2 \le N_c \le 5$.

\begin{table}
\caption{\footnotesize{Values of the IR zero of the beta function in the
supersymmetric SU($N_c$) gauge theory with $N_f$ pairs of chiral superfields in
$\Phi_i$, $\tilde\Phi_i$ in 
the fundamental and conjugate fundamental represention,
respectively, calculated at $n$-loop order, and denoted
as $\alpha_{IR,n\ell}$.  Results are given for the illustrative values 
$2 \le N_c \le 5$. For each $N_c$, we only give results for the integral
$N_f$ values in the interval (\ref{nfinterval}) where the theory is
asymptotically free and the two-loop beta function has an infrared zero.}}
\begin{center}
\begin{tabular}{|c|c|c|c|} \hline\hline
$N_c$ & $N_f$& $\alpha_{IR,2\ell}$ & $\alpha_{IR,3\ell}$ \\ \hline
 2    & 4   & 6.28                & 2.65   \\
 2    & 5   & 1.14                & 0.898  \\
 \hline
 3    & 5   & 18.85               & 3.05   \\
 3    & 6   & 2.69                & 1.40   \\
 3    & 7   & 0.992               & 0.734  \\
 3    & 8   & 0.343               & 0.308  \\
\hline
 4    & 7   & 5.03                & 1.64   \\
 4    & 8   & 1.795               & 0.984  \\
 4    & 9   & 0.867               & 0.615  \\
 4    & 10  & 0.426               & 0.357  \\
 4    & 11  & 0.169               & 0.158  \\
\hline
 5    & 8   &12.94                & 1.90   \\
 5    & 9   & 2.86                & 1.13   \\
 5    & 10  & 1.37                & 0.765  \\
 5    & 11  & 0.766               & 0.528  \\
 5    & 12  & 0.442               & 0.353  \\
 5    & 13  & 0.240               & 0.212  \\
 5    & 14  & 0.101               & 0.0963 \\
\hline\hline
\end{tabular}
\end{center}
\label{betazero_fund}
\end{table}

It is often of interest to consider the 't Hooft limit $N_c \to \infty$ with
$g^2N_c$ fixed and finite.  For the present $F+\bar F$ superfield content
it is also natural to consider taking $N_f \to \infty$ with the ratio
\beq
r \equiv \frac{N_f}{N_c} 
\label{r}
\eeq
fixed and finite (sometimes called the Veneziano limit).  In this limit, the
relevant interval for $r$ where the two-loop beta function has an IR zero is 
thus
\beq
\frac{3}{2} < r < 3 \ . 
\label{rinterval}
\eeq
Here, 
\beq
\alpha_{IR,2\ell} N_c = \frac{2\pi(3-r)}{2r-3} \ , 
\label{alfir2loop_red}
\eeq
which decreases monotonically to 0 as $r$ increases through the interval 
$3/2 < r < 3$.

\subsubsection{Three-Loop Analysis}

For the present case, Eq. (\ref{fundrep}), the general result in Eq. (\ref{b3})
for the three-loop coefficient $b_3$ takes the form
\beqs
b_3 & = & 21N_c^3 + (9-21N_c^2+2N_c^{-2})N_f \cr\cr
    & + & (4N_c - 3N_c^{-1})N_f^2 \ . 
\label{b3_fund}
\eeqs
For small $N_f$, $b_3$ is positive.  As $N_f$ increases, $b_3$ passes through
zero and reverses sign from positive to negative.  To investigate this, one
solves the equation $b_3=0$ for $N_f$.  Since $b_3$ is a quadratic function of
$N_f$, there are formally two solutions to this equation, namely
\beqs
N_{f,b3z,\pm} & = & [2N_c(4N_c^2-3)]^{-1} \bigg [ 21N_c^4-9N_c^2-2 \cr\cr
& \pm & \sqrt{105N_c^8-126N_c^6-3N_c^4+36N_c^2+4} \ \bigg ]  \ . \cr\cr
& & 
\label{nfb3z}
\eeqs
For $N_c=2$, this gives $N_{f,b3z,-} = 3.09$ (to the indicated accuracy),
slightly above $(3/2)N_c=3$, while for $N_c \ge 3$, we find that $N_{f,b3z} <
(3/2)N_c$. As examples, for $N_c=3$, $N_{f,b3z,-} = 4.27 < (3/2)N_c = 4.5$,
while for $N_c=4$, $N_{f,b3z,-} = 5.55 < (3/2)N_c = 6$, and so forth for higher
values of $N_c$. For large $N_c$, 
\beq
\frac{N_{f,b3z,-}}{(3/2)N_c} = \frac{1}{12}\Big [ 21-\sqrt{105} \ \Big ] 
+ O \bigg ( \frac{1}{N_c^2} \bigg ) \ . 
\label{nfb3z_minus_taylor}
\eeq
The numerical value of the first term is approximately 0.896. As $N_f$
increases past the larger value $N_{f,b3z,+}$, $b_3$ vanishes and reverses sign
again, becoming positive. However, this larger zero is not
relevant to our analysis, since for all $N_c \ge 2$, $N_{f,b3z,+} > 3N_c$,
i.e., this occurs for $N_f$ beyond the upper limit imposed by the constraint of
asymptotic freedom.  For example, for $N_c=2$,
$N_{f,b3z,+} \simeq 8.38$, which is greater than $N_{f,max} = 6$, and so 
forth for larger values of $N_c$. For large $N_c$, we have
\beq
\frac{N_{f,b3z,+}}{3N_c} = \frac{1}{24}\Big [21+\sqrt{105} \ \Big ]
+ O \bigg ( \frac{1}{N_c^2} \bigg ) \ . 
\label{nfb3z_minus_taylor2}
\eeq
The numerical value of the first term is 1.30.  

Since we restrict here to the interval (\ref{nfinterval}) where the two-loop
beta function has an IR zero, we observe that for physical, integral values of
$N_f$ in this interval, $b_3$ is always negative.  Note that for $N_c=2$, $b_3$
vanishes and changes sign from positive to negative as $N_f$ increases through
$N_f=3.09$, but this value of $N_f$ is less than the value $N_{f,b2z} = 3.43$
where the beta function first has an IR zero for this $N_c$.  Hence, for all
$N_c \ge 2$ and for integral values of $N_f$ in the interval (\ref{nfinterval})
where there is an IR zero at the two-loop level, $b_3 < 0$.

We list values of $\alpha_{IR,3\ell}$ for $2 \le N_c \le 5$ in 
Table \ref{betazero_fund}, together with the values of 
$\alpha_{IR,2\ell}$ already given. As is evident from this table, 
\beq
\alpha_{IR,3\ell} < \alpha_{IR,2\ell} \ . 
\eeq
This is the same trend that we found in \cite{bvh} for a non-supersymmetric
SU($N_c$) with $N_f$ copies of massless fermions in the fundamental
representation.  

In the large-$N_c$, large-$N_f$ limit with $N_f=rN_c$, we calculate 
\beq 
\alpha_{IR,3\ell} N_c = \frac{4\pi\Big [ 3-2r + \sqrt{4r^3-29r^2+72r-54} \
\Big ]} {21(r-1)-4r^2} \ . 
\label{alfir3loop_red}
\eeq
The right-hand side of Eq. (\ref{alfir3loop_red}) decreases monotonically from
$4\pi \simeq 12.57$ to 0 as $r$ increases from 3/2 to 3. Note that the
denominator in Eq. (\ref{alfir3loop_red}), $21(r-1)-4r^2$, is positive-definite
in the relevant interval $3/2 < r < 3$; it has zeros at $r=(1/8)(21 \pm
\sqrt{105})$, i.e., approximately 3.906 and 1.344.  In the numerator, the first
part, $3-2r$, is negative-definite in this interval, but is smaller than the
square root.

% =========================================================================

\subsection{Values of $\gamma_m$ at IR Zero of $\beta$}

\subsubsection{Coefficients} 

Evaluating Eqs. (\ref{c1})-(\ref{c3}) for our case (\ref{fundrep}), we have
\beq
c_1 = \frac{2(N_c^2-1)}{N_c} \ ,  
\label{c1fund}
\eeq
\beq
c_2 = 2(N_c^2-1)N_c^{-2}(2N_c^2+1-N_cN_f) \ , 
\label{c2_fund}
\eeq
and
\begin{widetext}
 \beq
c_3 =  2(N_c^2-1)N_c^{-3}\bigg [ 5N_c^4-N_c^2+2+N_c(-3N_c^2+4)N_f 
-6\zeta(3)N_c(N_c^2+1)N_f-N_c^2N_f^2 \bigg ]  \ , 
\label{c3_fund}
\eeq
\end{widetext}
where $\zeta(s)$ is the Riemann zeta function.  As $N_f$ approaches
$N_{f,max}$ from below, $b_1 \to 0$ with nonzero $b_2$ and hence
$\alpha_{IR} \to 0$; since the perturbative calculation expresses $\gamma_m$ in
a power series in $\alpha$, it follows that as $\gamma_m \to 0$ as $N_f \to
N_{f,max}$.

To get an analytic understanding of the behavior of $\gamma_m$, we study the
signs of the coefficients $c_\ell$ with $\ell=1, \ 2, \ 3$.  Since the one-loop
coefficient $c_1$ is positive, $\gamma_m$ increases from zero as $N_f$
decreases just below $N_{f,max}$.  In the conformal phase the two-loop
coefficient $c_2$ may be either positive or negative, depending on $N_f$.  This
coefficient $c_2$ vanishes at 
\beq
N_f = 2N_c + N_{c}^{-1} \equiv N_{f,c2z} \ . 
\label{nfc2z}
\eeq
and
\beqs
c_2 & > & 0 \quad {\rm for} \ \ N_f < N_{f,c2z} \ , \cr\cr
c_2 & < & 0 \quad {\rm for} \ \ N_{f,c2z} < N_f < N_{f,max} = 3N_c \ . 
\label{c2signs}
\eeqs

The three-loop coefficient $c_3$ is a quadratic function of $N_f$ and
vanishes at two values of $N_f=0$, namely 
\beq
N_{f,c3z,fund,\pm}=\frac{4-3N_c^2-6(N_c^2+1)\zeta(3) \pm \sqrt{R_{c3z}}}{2N_c}
\label{nfc3z_fund}
\eeq
where
\beqs
R_{c3z} & = & 29N_c^4-28N_c^2+24+12(N_c^2+1)(3N_c^2-4)\zeta(3) \cr\cr
        & + & [6(N_c^2+1)]^2 \zeta(3)^2 \ . 
\label{rc3z}
\eeqs
For $N_c=2$, $N_{f,c3z,fund,-}$ is equal to $-22.88$ and hence is unphysical,
while $N_{f,c3z,fund,+} = 0.8522$, so that in the physical range, $c_3$ is
negative for all positive integral values of $N_f$.  For $N_c=3$,
$N_{f,c3z,fund,-}$ is equal to $-33.05$ and hence is again unphysical, while
$N_{f,c3z,fund,+} = 1.338$, so that in the physical range, $c_3$ is positive
for $N_f =1$, but negative for $N_f \ge 2$, including all of the interval of
interest here.  This qualitative behavior continues to hold for higher values
of $N_c$, as is evident from the Taylor series expansion
\beqs
& & \frac{N_{f,c3z,fund,+}}{N_c} = \frac{1}{2} \bigg [ -3(1+2\zeta(3)) \cr\cr
& + & \sqrt{36\zeta(3)(\zeta(3)+1) + 29} \ \bigg ] + O \bigg ( \frac{1}{N_c^2} 
\bigg ) \ . 
\label{nf3czplus_fund_taylor}
\eeqs
The numerical value of the constant term is 0.468 to the indicated accuracy.
Thus, as $N_c \to \infty$, $N_{f,c3z,fund,+} \sim 0.468 N_c$, which is less
than the exact value of the lower boundary of the conformal phase,
$N_{f,cr}=(3/2)N_c$, given in Eq. (\ref{nfcr_fund}). Thus, a general
characterization of the three-loop $\gamma_m$ is that $c_1$ is positive for any
representation $R$, and for $R=F$; $c_2$ is positive in the lower part of the
conformal phase but negative in the upper part, as specified by
Eq. (\ref{c2signs}), and $c_3$ is negative throughout all of the $N_f$ interval
of interest, including the conformal phase.

\subsubsection{$\gamma_{IR,SI}$ }

\begin{figure}
  \begin{center}
    \includegraphics[width=8.2cm]{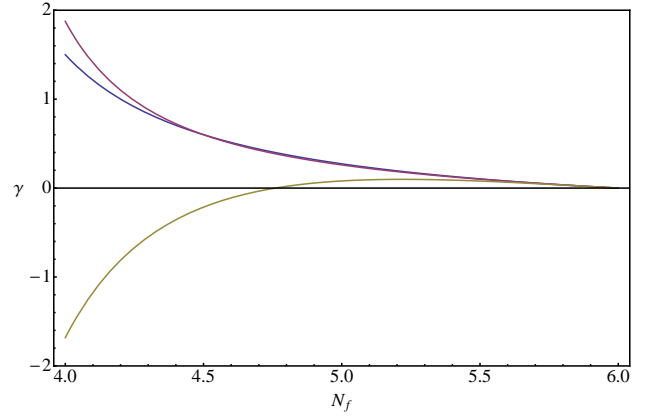}
  \end{center}
\caption{\footnotesize{ Plot of the $n$-loop fermion anomalous dimension,
$\gamma_{n\ell}$, evaluated at the respective $n$-loop value of
the IR zero of $\beta$, $\alpha_{IR,n\ell}$, and denoted as
$\gamma_{IR,n\ell}$, for two-loop and three-loop order, in the case of $N_f$
chiral superfields $\Phi_i, \ \tilde\Phi_i$ in the $F$ and $\bar F$
representation of SU($N_c$) for $N_c=2$.  We use the generic label $\gamma$ for
the vertical axis. At the lower end of the plot, from
top to bottom, the curves are for (i) $\gamma_{IR,2\ell}$, (ii)
$\gamma_{IR,SI}$, and (iii) $\gamma_{IR,3\ell}$.  The curves involve an
implicit analytic continuation of $N_f$ from integer values to real values; of
course, only the integer values are physical. We only show the region in $N_f$
where $\gamma_{IR,2\ell}$ and $\gamma_{IR,SI}$ approximately satisfy the upper
bound (\ref{gammabound}).}}
\label{susygamma_fund_nc2}
\end{figure}

\begin{figure}
  \begin{center}
    \includegraphics[width=8.2cm]{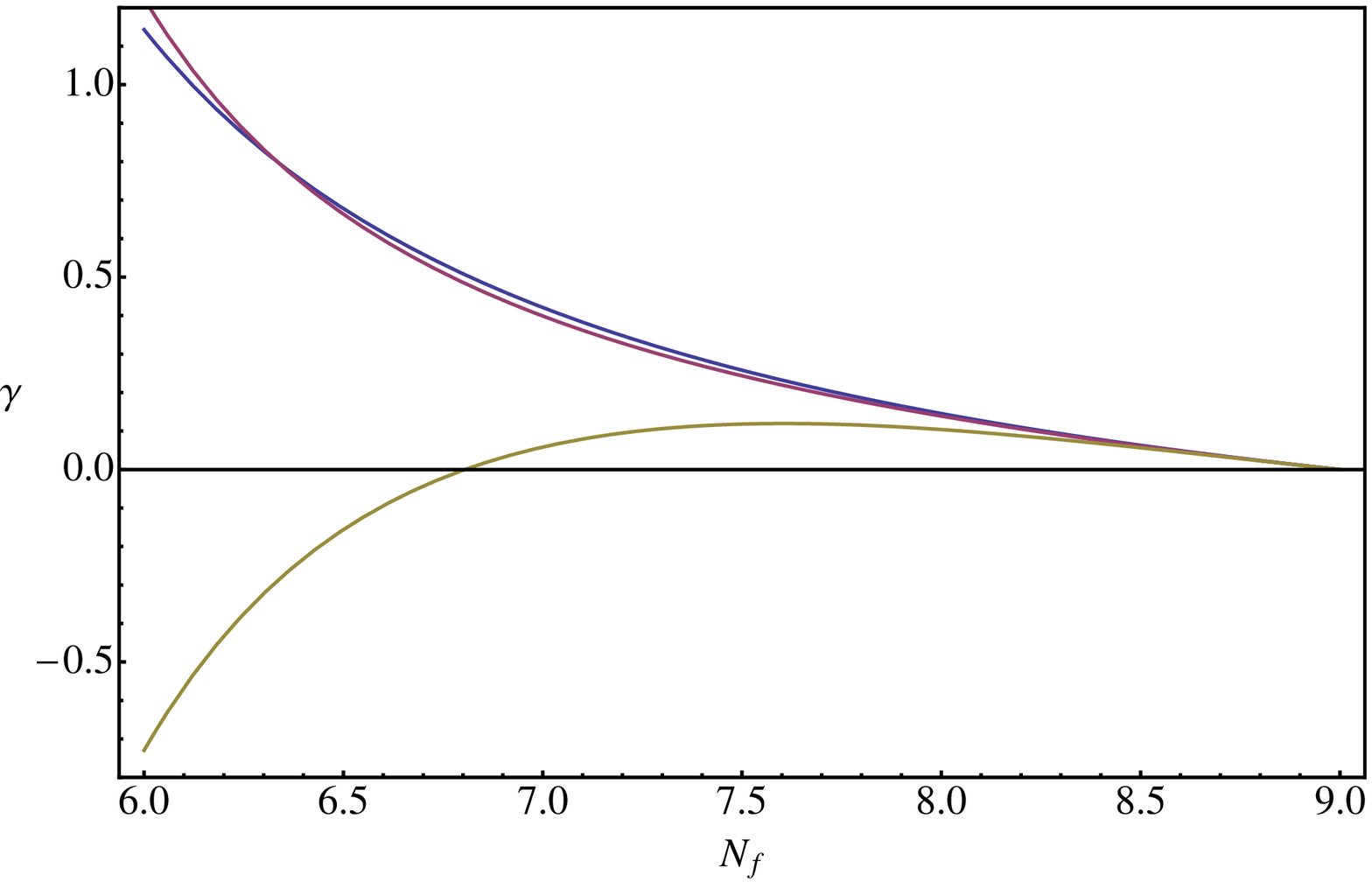}
  \end{center}
\caption{\footnotesize{Same as Fig. \ref{susygamma_fund_nc2} for $N_c=3$.}}
\label{susygamma_fund_nc3}
\end{figure}

\begin{figure}
  \begin{center}
    \includegraphics[width=8.2cm]{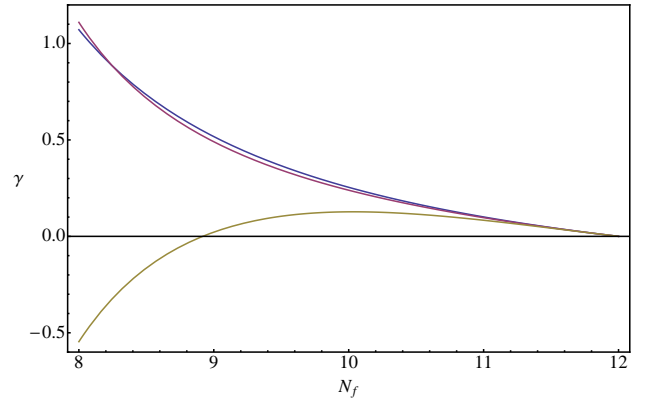}
  \end{center}
\caption{\footnotesize{Same as Fig. \ref{susygamma_fund_nc2} for $N_c=4$.}}
\label{susygamma_fund_nc4}
\end{figure}

For the chiral fermion content (\ref{fundrep}), we calculate 
\beq
\gamma_{IR,SI} = \frac{(N_c^2-1)(3N_c-N_f)}{(2N_c^2-1)N_f-3N_c^3} \ . 
\label{gammasi_fund}
\eeq
As $N_f$ decreases below its maximal value, $3N_c$, $\gamma_{IR,SI}$ increases
monotonically.  We list values of $\gamma_{IR,SI}$ in Table 
\ref{gammavalues_fund}
for the illustrative values $2 \le N_c \le 5$.
\begin{table}
\caption{\footnotesize{Values of the anomalous dimension $\gamma_m$ in the
SU($N_c$) supersymmetric gauge theory with $N_f$ copies of massless chiral
superfields $\Phi_i$, $\tilde \Phi_i$ in the $F$ and $\bar F$ representations,
calculated to the $n$-loop order in perturbation theory and evaluated at the IR
zero of the beta function calculated to this order, $\alpha_{IR,n\ell}$, for
$\ell=2,3$.  We denote these as $\gamma_{IR,n\ell} \equiv
\gamma_{n\ell}(\alpha_{IR,n\ell})$.  Results are given for the illustrative
values $2 \le N_c \le 5$. For sufficiently small $N_f > N_{f,b2z}$ for each
$N_c$, $\alpha_{IR,2\ell}$ is so large that the formal values of
$\gamma_{IR,2\ell}$ and/or $\gamma_{IR,3\ell}$
are either larger than unity or negative and hence are unphysical. We
indicate this by placing these values in parentheses.}}
\begin{center}
\begin{tabular}{|c|c|c|c|c|} \hline\hline
$N_c$ & $N_f$& $\gamma_{IR,SI}$ & $\gamma_{IR,2\ell}$ & $\gamma_{IR,3\ell}$ 
\\ \hline
 2    & 4    & (1.500)  & (1.875)  & $(-1.68)$     \\
 2    & 5    & 0.273    & 0.260    &    0.0802     \\
 \hline
 3    & 6    & (1.14)   & (1.22)   & $(-0.730)$    \\
 3    & 7    & 0.421    & 0.399    & 0.0584        \\
 3    & 8    & 0.145    & 0.139    & 0.104         \\
\hline
 4    & 8    & (1.07)   & (1.11)   & $(-0.546)$    \\
 4    & 9    & 0.517    & 0.490    & 0.0219        \\
 4    & 10   & 0.254    & 0.239    & 0.127         \\
 4    & 11   & 0.101    & 0.0970   & 0.0835        \\
\hline
 5    & 10   & (1.04)   & (1.07)   & $(-0.475)$    \\
 5    & 11   & 0.585    & 0.557    & $(-0.0135)$   \\
 5    & 12   & 0.338    & 0.317    & 0.120         \\
 5    & 13   & 0.183    & 0.173    & 0.121         \\
 5    & 14   & 0.0772   & 0.0748   & 0.0680        \\
\hline\hline
\end{tabular}
\end{center}
\label{gammavalues_fund} 
\end{table}
For each $N_c$, we omit values
in the lower range of $N_f$ that strongly violate the bound (\ref{gammabound}).
This violation is due to both the inexactness of the perturbative calculation
of $\gamma_m$ and the fact that the two-loop IR zero of the beta function,
$\alpha_{IR,2\ell}$ gets arbitrarily large as $N_f$ decreases toward
$N_{f,b2z}$. In Figs. \ref{susygamma_fund_nc2}-\ref{susygamma_fund_nc4} we show
plots of $\gamma_{IR,SI}$ as a function of $N_f$ for the approximate respective
subintervals of the conformal phase where $\gamma_{IR,SI}$ satisfies the upper
bound (\ref{gammabound}). We will discuss below the other curves on these
plots.

Specializing Eq. (\ref{nfcrsi}) to the case of the fundamental representation,
we obtain the scheme-independent perturbative estimate of the lower boundary 
of the conformal phase, 
\beq
N_{f,cr,SI} = \frac{3N_c(2N_c^2-1)}{3N_c^2-2} \ . 
\label{nfcrsi_fund}
\eeq
As noted before, this estimate lies above the actual exact lower
boundary, which, in the present case, occurs at $N_{f,cr}=(3/2)N_c$:
\beq
\frac{N_{f,cr,SI}}{N_{f,cr}} = \frac{2(2N_c^2-1)}{3N_c^2-2} \ . 
\label{nfcrsiratio_fund}
\eeq
This ratio decreases from the value $7/5=1.40$ at $N_c=2$ to 4/3 as $N_c \to 
\infty$ and has the Taylor series expansion
\beq
\frac{N_{f,cr,SI}}{N_{f,cr}} = \frac{4}{3} + \frac{2}{9N_c^2} + 
O \bigg ( \frac{1}{N_c^4} \bigg ) \quad {\rm as} \ \ N_c \to \infty \ . 
\label{nfcrsiratio_fund_taylor}
\eeq

We next proceed to two-loop and three-loop analyses.  As we have remarked
above, these have the advantage of using the same $n$-loop orders in
calculating $\gamma_m$ and $\alpha_{IR,n\ell}$, but are subject to the standard
caution that they involve scheme-dependence.

\subsubsection{Two-Loop Analysis}

Evaluating $\gamma_m$ calculated to two-loop order at the zero of the beta
function calculated to the same order for $R=F$, we obtain
\beqs
& & \gamma_{IR,2\ell} = 
\frac{(N_c^2-1)(3N_c-N_f)(N_f-N_c)(N_cN_f-3)}
{2(-3N_c^3+2N_c^2N_f-N_f)^2} \ . \cr\cr
& & 
\label{gam2loop_fund_irfp}
\eeqs
We list values of $\gamma_{IR,2\ell}$ in Table \ref{gammavalues_fund} and plot
curves of $\gamma_{IR,2\ell}$ as a function of $N_f$ (analytically continued
from integer to real values) in
Figs. \ref{susygamma_fund_nc2}-\ref{susygamma_fund_nc4}. One sees from these
results that over the range where the calculations are reliable,
$\gamma_{IR,2\ell}$ is rather close to $\gamma_{IR,SI}$.  The anomalous
dimension $\gamma_{IR,2\ell}$ increases monotonically as $N_f$ decreases from
$N_{f,max}=3N_c$ in the interval (\ref{nfinterval}).  As $N_f$ decreases toward
$N_{f,b2z}$ (given in Eq. (\ref{nfb2z_fund}), $\alpha_{IR,2\ell}$ gets
arbitrarily large, and this perturbative expression obviously ceases to be
reliable.  As $N_f$ increases toward $N_{f,max,fund}=3N_c$, $\alpha_{IR,2\ell}
\to 0$, and hence also $\gamma_{IR,2\ell} \to 0$.

The condition that $\gamma_{IR,2\ell} = 1$ is a cubic equation in $N_f$, 
namely 
\beqs
& & N_c(N_c^2-1)N_f^3 + (4N_c^4-7N_c^2+5)N_f^2 \cr\cr 
& + & 3N_c(-7N_c^4+7N_c^2-4)N_f  + 9N_c^2(2N_c^4-N_c^2+1) = 0 \ . \cr\cr
& & 
\label{gam2loopirfpeq1}
\eeqs
As $N_f$ decreases from $N_{f,max,fund}=3N_c$ toward $N_{f,b2z,fund}$, 
$\gamma_m$
exceeds the rigorous upper bound $\gamma_m < 1$ at a value of $N_f$ which is
given as the relevant one among the three roots of Eq. (\ref{gam2loopirfpeq1}).
For example, for $N_c=2$, this equation has the physical root $N_f = 4.242$,
together with two other roots which are not relevant, namely $N_f = 2.929$ and
$N_f=-14.00$ (to four significant figures).  The first of these other roots is
irrelevant since it is below the lower boundary of the IR conformal phase,
$N_{f,cr}$ given in Eq. (\ref{nfcr_fund}), and the second is irrelevant since
it is negative and hence unphysical. Thus, insofar as one can compare a
perturbative 2-loop calculation of $\gamma_m$ with an upper bound on the exact
$\gamma_m$, one finds that this 2-loop prediction for the lower end of the
conformal phase in $N_f$ for $N_c=2$ is $N_f = 4.242$.  The ratio of this to
the exact result $N_{f,cr}$, which is 3 for $N_c=2$, is approximately 1.414.
Similarly, for $N_c=3$, the physical root of Eq. (\ref{gam2loopirfpeq1}) is
$N_f=6.150$. (The other two roots are 3.983, and $-21.22$, which are again
irrelevant).  The ratio of this to the (formal, half-integral) exact result
$N_{f,cr}=9/2$ is 1.367.  As $N_c \to \infty$, we find that this ratio of the
physical root of the cubic for $N_{f,cr}$ divided by the exact value
$N_{f,cr}=(3/2)N_c$ approaches the same value, 4/3, as was true of the ratio of
$N_{f,cr,SI}$ divided by this exact value, as given in 
Eqs. (\ref{nfcrsiratio_fund}) and (\ref{nfcrsiratio_fund_taylor}) above:
\beq
\frac{N_{f,cr,\gamma_{IR,2\ell}}}{N_{f,cr}} = \frac{4}{3} + O\bigg ( 
\frac{1}{N_c^2} \bigg ) \ . 
\label{nfcrratiolargenc}
\eeq
We thus find that a 2-loop perturbative analysis of $\beta$ and $\gamma_m$
overestimates the value of $N_{f,cr}$ somewhat and hence underestimates the
size of the conformal phase in $N_f$ for this case (\ref{fundrep}).  This is
qualitatively the same as we found for the estimate of $N_{f,cr}$ using
$\gamma_{IR,SI}$.

It is interesting to compare these estimates for $N_{f,cr}$ from the
scheme-independent $\gamma_{IR,SI}$ and the two-loop $\gamma_{IR,2\ell}$ with
the results obtained via the different method of equating $\alpha_{IR,2\ell}$
with the critical value $\alpha_{cr}$ for dynamical mass generation calculated
from the Dyson-Schwinger equation for the fermion propagator in
Ref. \cite{ans98}.  This mass generation, associated with fermion
condensation, was found to occur at \cite{ans98}
\beq
N_{f,cr,DS} = \frac{3N_c(3N_c^2-2)}{4N_c^2-3} \ . 
\label{nfans}
\eeq
The ratio of this to the exact value for $N_{f,cr}$ in Eq. (\ref{nfcr_fund}) is
\beq
\frac{N_{f,cr,DS}}{N_{f,cr}} = \frac{3}{2} \bigg ( \frac{1-(2/3)N_c^{-2}}
{1-(3/4)N_c^{-2}} \bigg ) \ ,
\label{nfcrdsratio}
\eeq
This ratio is equal to 1.54 and 1.515 for $N_c=2$ and $N_c=3$ and decreases
monotonically to 3/2 as $N_c \to \infty$.  Thus, for a given $N_c$, our
estimates of $N_{f,cr}$ obtained from equating $\gamma_{IR,SI}$ and
$\gamma_{IR,2\ell}$ to the upper bound of unity are slightly closer to the
exact result (\ref{nfcr_fund}) than the estimate from the analysis of the
Dyson-Schwinger equation, and all agree qualitatively, i.e., all are slight
overestimates of $N_{f,cr}$.

In the large-$N_c$, large-$N_f$ limit, with $N_f=rN_c$, $\gamma_{IR,2\ell}$ 
has the Taylor series expansion 
\beqs
& & \gamma_{IR,2\ell} = \frac{r(r-1)(3-r)}{2(3-2r)^2} -
\frac{3(r-1)(3-r)^2}{2(3-2r)^3N_c^2} + O \bigg ( \frac{1}{N_c^4} \bigg ) \ .
\cr\cr
& & 
\label{gam2loop_fund_irfp_largenc}
\eeqs
In this limit, the value of $r$, and hence $N_f$, where $\gamma_{IR,2\ell}=1$
is given as the relevant one among the three roots of the cubic equation
$(r-2)(r^2-6r-9)=0$, namely $r=2$.  The other two roots are $-3 + 3\sqrt{2}$
and $-3 - 3\sqrt{2}$; the first of these has the value 1.243 and hence is below
the value $r=3/2$ (cf. Eq. (\ref{nfb2z_fund})) where, for increasing $r$, an IR
zero of the beta function first appears (which coincides with the exact value
of the lower boundary of the IR conformal phase in this limit), and the second
is negative and hence obviously unphysical. Thus, in this large-$N_c$ limit,
with $r=N_f/N_c$ fixed and finite, the 2-loop analysis predicts that
$N_{f,cr}=2N_c$, which is 4/3 times the exact result of 
Eq. (\ref{nfcr_fund}), in agreement with our analysis in Eq. 
(\ref{nfcrratiolargenc}).

\subsubsection{Three-Loop Analysis}

In the same manner, we evaluate the three-loop expression for $\gamma_m$ at the
3-loop value of the IR zero of the beta function $\alpha_{IR,3\ell}$.  Since
the analytic formulas are somewhat complicated, we will restrict ourselves to
giving numerical results for $N_c=2$ through $N_c=4$ in Table
\ref{gammavalues_fund} and Figs.
\ref{susygamma_fund_nc2}-\ref{susygamma_fund_nc4}.  In contrast to
$\gamma_{IR,SI}$ and $\gamma_{IR,2\ell}$, we find that $\gamma_{IR,3\ell}$ does
not increase monotonically as $N_f$ decreases below $N_{f,max}$. Instead, it
reaches a maximum well below unity in the interior of the conformal phase and
then decreases, vanishing and becoming negative.  Because of this behavior, we
cannot use our procedure of setting the perturbative expression for $\gamma_m$
equal to 1 and then solving for $N_f$ for $\gamma_{IR,3\ell}$.  Since this
behavior of $\gamma_{IR,3\ell}$ clearly differs from the behavior of the
scheme-independent $\gamma_{IR,SI}$, it may reflect scheme-dependence.  It also
shows that as $N_f$ decreases toward the lower end of the IR conformal phase
and $\alpha_{IR}$ increases to larger values, a perturbative calculation
becomes less reliable.  In general, a necessary condition for these
perturbative calculations to be reliable is that inclusion of the next
higher-loop order term should not drastically change the qualitative behavior.
Imposing this condition for the comparison of $\gamma_{IR,2\ell}$ and
$\gamma_{IR,3\ell}$, we may obtain an estimate of the interval in $N_f$ where
the calculation could be reasonably reliable.  For $N_c=2$, we find that this
interval plausibly includes $N_f=5$ but does not include $N_f=4$.  For $N_c=3$,
this interval includes $N_f=8$ but not lower values of $N_f$.  In view of this
behavior of $\gamma_{IR,3\ell}$, one must view the three-loop results with 
appropriate caution, recognizing that perturbative calculations become less 
reliable as the coupling becomes stronger.  

% ========================================================================

\section{Superfields in the Adjoint Representation}

\subsection{Beta Function}

The adjoint representation is self-conjugate, so here a theory with $N_f$
copies of a massless chiral superfield content consisting of $\Phi_i$ and
$\tilde \Phi_i$, $i=1,...,N_f$, is equivalent to a theory with $N_f'=2N_f$
copies of $\Phi_i$. We shall thus consider half-integral values of $N_f$ as
physical here.  The beta function coefficients are
\beq
b_1 = N_c(3-2N_f) \ , 
\label{b1adj}
\eeq
\beq
b_2 = -6N_c^2(2N_f-1) 
\label{b2adj}
\eeq
and, in the $\overline{DR}$ scheme, 
\beq
b_3 = -7N_c^3(2N_f-1)(3-2N_f) \ . 
\label{b3adj}
\eeq
Note that $b_3$ vanishes at the same (formal, non-integral) value of $N_f$ at
which $b_1$ vanishes, namely $N_f=3/2$.  The condition that the theory be
asymptotically free, i.e., that $b_1 > 0$, implies the upper bound $N_f < 3/2$.
\beq
N_f < \frac{3}{2} = N_{f,max} 
\label{nfmaxadj}
\eeq
For the $\Phi_i$, $\tilde \Phi_i$ content, this only allows the choice $N_f=1$,
while for the reduced content consisting only of $\Phi_i$, this allows $N_f'=1$
and $N_f'=2$.  (Note that the $N_f'=1$ theory has been solved exactly
\cite{sw94}.) 

From Eq. (\ref{nfcr}), the lower boundary of the IR conformal phase is given
formally by
\beq
N_{f,cr} = \frac{3}{4}
\label{nfcr_adj}
\eeq
or equivalently, $N'_{f,cr}=3/2$. Since neither of these is an integer, they
must be regarded only as quantities defined via a requisite analytic
continuation of the theory in $N_f$ or $N_f'$ away from the integers to the
real numbers.  With this understanding, the IR conformal phase is thus given by
\beq
\frac{3}{4} < N_f < \frac{3}{2} \ . 
\label{nfconformaladj}
\eeq
Hence, with the $\Phi_i$, $\tilde \Phi_i$ superfield content, 
the only integer value of $N_f$ allowed by the requirement of 
asymptotic freedom, namely $N_f=1$ yields an IR conformal phase.  For the
theory with just the $\Phi_i$ superfield, Eq. (\ref{nfconformaladj}) reads 
$3/2 < N_f'< 3$, so for $N_f'=2$ ($N_f'=1$) the theory evolves into the
infrared in a conformal (nonconformal) manner, respectively. 

In the theory with $\Phi$, $\tilde \Phi$ superfield content, the two-loop
$\beta$ function coefficient $b_2$ is negative for the only relevant value of
$N_f$, namely $N_f=1$.  In the reduced theory, $b_2=-6N_c^2(N_f'-1)$, which is
zero for $N_f'=1$ and negative for $N_f'=2$. Thus, at the two-loop level, the
IR zero of the $\beta$ functions occurs at
\beq
\alpha_{IR,2\ell} = \frac{2\pi(3-2N_f)}{3N_c(2N_f-1)} \ . 
\label{alfir2loop_adj}
\eeq

At the three-loop level, $\beta$ has two zeros away from the origin, at
\beq
\frac{\alpha}{4\pi}
 = \frac{-3 \pm \sqrt{ \frac{2(14N_f^2-33N_f+27)}{2N_f-1}}}{7N_c(3-2N_f)}
\ .
\label{alfir3loopadj}
\eeq
The $-$ sign choice yields an unphysical, negative result, so the 
physical three-loop IR zero of the beta function is given by
Eq. (\ref{alfir3loopadj}) with the $+$ sign.  We denote this as
$\alpha_{IR,3\ell}$ (suppressing the $Adj$ for simplicity). This
$\alpha_{IR,3\ell}$ exhibits unphysical behavior, vanishing at $N_f=9/8$ and 
becoming negative in the range $9/8 < N_f < N_{f,max}=3/2$. One could take the
point of view that this precludes a reliable three-loop perturbative analysis 
of this case.  However, we will at least give results for the one case for
which the theory has an IR fixed point, namely $N_f=1$. For $N_f=1$, we have 
\beq
\alpha_{IR,2\ell,N_f=1} = \frac{2\pi}{3N_c}
\label{alfir2loop_adj_nf1}
\eeq
and
\beq
\alpha_{IR,3\ell,N_f=1} = \frac{4\pi}{7N_c} \ , 
\label{alfir3loop_adj_nf1}
\eeq
so that, for this value of $N_f$,
\beq
\frac{\alpha_{IR,3\ell}}{\alpha_{IR,2\ell}} = \frac{6}{7} \ , 
\label{alfir3loopover2loop_adj}
\eeq
independent of $N_c$. This is the same trend that we found for the case of
matter superfields in the $F + \bar F$ representation, i.e., the value of the
IR fixed point calculated to three-loop order is somewhat smaller than the
value calculated to two-loop order.

\subsection{Anomalous Dimension}

For this theory with chiral superfields in the adjoint representation, the
coefficients in Eq. (\ref{gamma}) are
\beq
c_1 = 4N_c \ , 
\label{c1_adj}
\eeq
\beq
c_2 = -4N_c^2(2N_f-1) \ , 
\label{c2_adj}
\eeq
and
\beq
c_3 = -8N_c^3(N_f+4)(2N_f-1) \ .
\label{c3_adj}
\eeq
From our general result (\ref{gammasi}), we calculate the scheme-independent
anomalous dimension  
\beq
\gamma_{IR,SI} = \frac{2(3-2N_f)}{3(2N_f-1)} \ . 
\label{gammasi_adj}
\eeq
This increases monotonically as $N_f$ decreases from its maximal to its minimal
value in the IR conformal phase.  $\gamma_{IR,SI}$ increases through its upper
limit of unity as $N_f$ decreases through the value $N_{f,cr,SI}$.  Evaluating
our general formula in Eq. (\ref{nfcrsi}) for the present case of the adjoint
representation, we obtain 
\beq
N_{f,cr,SI} = \frac{9}{10} \ .
\label{nfcrsi_adj}
\eeq
As is true for general $R$, this is larger than the exact result, which in the
present case is $N_{f,cr}=3/4$.  (This exact value is only formal, since it is
non-integral.)  The ratio (\ref{nfcrsiratio}) here is 
6/5 = 1.2.  As $N_f$ decreases from 9/10 to $N_{f,cr}=3/4$, $\gamma_{IR,SI}$
increases from 1 to 2, exhibiting unphysical behavior.  

Evaluating Eq. (\ref{gammair2loop}) for the present case of the adjoint 
representation, we find 
\beq
\gamma_{IR,2\ell} = \frac{(3-2N_f)(2N_f+3)}{9(2N_f-1)} \ . 
\label{gamma_2loop_adj}
\eeq
Setting this equal to 1, we derive another perturbative estimate of $N_{f,cr}$,
which is the positive root of a quadratic equation, 
\beq
N_f = \frac{3(-3+\sqrt{17} \ )}{4} \simeq 0.8423 \ .
\label{nfcr2loopadj}
\eeq
(The other root is negative).  This is again slightly larger than the formal
exact $N_{f,cr} = 3/4$. 

For the only physical value where there is an IR fixed point, $N_f=1$, we thus
have
\beqs
\gamma_{IR,SI}    & = & \frac{2}{3} = 0.6666... \cr\cr
\gamma_{IR,2\ell} & = & \frac{5}{9} = 0.5555... \cr\cr 
\gamma_{IR,3\ell} & = & \frac{2^7}{7^3} = 0.37317... \quad {\rm for} \ \ N_f=1 
\ . 
\label{gammavalues_adj}
\eeqs
Again, we find the same trend as for $\Phi_i$, $\tilde \Phi_i$ in $F+\bar F$,
namely that the value of the anomalous dimension $\gamma_m$ evaluated at the IR
fixed point decreases somewhat when one goes from two-loop order (or the
scheme-independent result) to three-loop order.

% =======================================================================

\section{Chiral Superfields in the Symmetric or Antisymmetric 
Rank-2 Tensor Representation}

In this section we analyze the UV to IR evoluation of the supersymmetric
SU($N_c$) theory with $\Phi_i$, $\tilde \Phi_i$ in the $R$ and $\bar R$
representation, where $R$ is a symmetric or antisymmetric rank-2 tensor
representation, denoted $S_2$, \ $A_2$, respectively.  Since many formulas are
closely related to each other, it is convenient to treat these two cases
together, as the $T_2$ representation.  In each of the combined formulas
involving a $\pm$ or $\mp$ sign, the upper and lower signs apply to the $S_2$
and $A_2$ representations, respectively.  For $N_c=2$, the $S_2$ representation
is the adjoint representation, which has already been discussed.  Thus, for the
$S_2$ representation, the distinct cases begin with $N_c \ge 3$.  For the $A_2$
case, $N_c$ is implicitly taken to be $N_c \ge 3$, since this representation is
the singlet if $N_c=2$. Further, note that the $A_2$ representation with
$N_c=3$ is equivalent to the conjugate fundamental representation so, with our
vectorlike content of chiral superfields, this reduces to the case $\Phi_i$,
$\tilde \Phi_i$ in the $F+\bar F$ representation already covered above.  Thus,
the $A_2$ cases that are distinct have $N_c \ge 4$. We focus here on
$\gamma_{IR,SI}$ and $\gamma_{IR,2\ell}$.

\subsection{$\beta$ Function}

We have 
\beq
b_1 = 3N_c -(N_c \pm 2)N_f \ , 
\label{b1t2}
\eeq
and
\beq
b_2 = 2 \Big [ 3N_c^2(1-N_f) \mp 8(N_c-N_c^{-1})N_f \Big ] \ . 
\label{b2t2}
\eeq
The expression for $b_3$ is similarly obtained in a straightforward manner from
the general result (\ref{b3}). The requirement of asymptotic freedom requires
$b_1 > 0$, i.e.,
\beq
N_f < N_{f,b1z} = N_{f,max} = \frac{3N_c}{N_c \pm 2} \ . 
\label{nfmax_t2}
\eeq
For the $S_2$ representation, $N_{f,max}$ increases monotonically from 3/2 for
$N_c=2$ to 3 as $N_c \to \infty$, while for the $A_2$ representation,
$N_{f,max}$ decreases monotonically from 9 for $N_c=3$ to 3 as $N_c \to
\infty$.

The exact result for the lower boundary of the IR conformal phase is
\beq
N_{f,cr} = \frac{N_{f,max}}{2} = \frac{3N_c}{2(N_c \pm 2)} \ . 
\label{nfcr_t2}
\eeq
For the $S_2$ representation, $N_{f,cr}$ increases monotonically from 3/4 for
$N_c=2$ to 3/2 as $N_c \to \infty$, while for the $A_2$ representation, 
$N_{f,cr}$ decreases monotonically from 9/2 for $N_c=3$ to 3/2 as 
$N_c \to \infty$.  Thus, the IR conformal phase exists for
\beq
\frac{3N_c}{2(N_c \pm 2)} < N_f < \frac{3N_c}{N_c \pm 2} \ . 
\label{irconformal_t2}
\eeq

The coefficient $b_2=0$ for 
\beq
N_f = N_{f,b2z} = \frac{3N_c^2}{3N_c^2 \pm 8(N_c-N_c^{-1})} \ . 
\label{nfb2z_t2}
\eeq
This is always smaller than $N_{f,cr}$ for the $S_2$ representation, so that
$b_2$ has fixed (negative) sign in the IR conformal phase in this case.  For
the $A_2$ representation, if $N_f < 1+\sqrt{5}= 3.236..$, then $N_{f,b2z} >
N_{f,cr}$, while if $3.22 < N_f < N_{f,max}$, then $N_{f,b2z} < N_{f,cr}$.
Hence, the only physical case where $N_{b2z} > N_{f,cr}$ is for the integer
value $N_c=3$, where the $A_2 + \bar A_2$ representation is equivalent to
the $F + \bar F$ representation.

At the two-loop level, the IR zero of $\beta$ occurs at $a_{IR,2\ell}=-b_1/b2$,
i.e., 
\beq
\alpha_{IR,2\ell} = \frac{2\pi[3N_c-(N_c \pm 2)N_f]}{3N_c^2(N_f-1)
\pm 8(N_c-N_c^{-1})N_f} \ . 
\label{alfir2loop_t2}
\eeq

\subsection{ $\gamma_m$}

For this $T_2$ case, 
\beq
c_1 = \frac{4(N_c \pm 2)(N_c \mp 1)}{N_c} \ , 
\label{c1_t2}
\eeq
and
\beq
c_2 = \frac{4(N_c \pm 2)(N_c \mp 1)[N_c^2 \mp 2N_c - 4 +N_c(N_c \pm 2)N_f]}
{N_c^2} \ . 
\label{c2_t2}
\eeq
The expression for $c_3$ is similarly obtained from the general result
(\ref{c3}). 

Hence,
\beq
\gamma_{IR,SI} = \frac{2(N_c \pm 2)(N_c \mp 1)[3N_c - (N_c \pm 2)N_f]}
{3N_c^3(N_f-1)\pm 8(N_c^2-1)N_c} \ .
\label{gammairsi_t2}
\eeq
and
\begin{widetext}
\beq
\gamma_{IR,2\ell} = \frac{(N_c \pm 2)(N_c \mp 1)[3N_c - (N_c \pm 2)N_f]
[N_c^2(3+N_f) \pm 2N_c(3+N_f) - 12][N_c(N_f-1) \pm 2N_f]}
{[3N_c^3(N_f-1)\pm 8(N_c^2-1)N_f]^2} \ . 
\eeq
\end{widetext}

One perturbative estimate of $N_{f,cr}$ is obtained by setting 
$\gamma_{IR,SI}=1$ and solving for $N_f$.  This gives
\beq
N_{f,cr,SI} = \frac{3N_c(3N_c^2\pm 2N_c-4)}{(N_c \pm 2)(5N_c^2\pm 4N_c-8)} \ . 
\label{nfcrgammasi_t2}
\eeq
Comparing these with the respective exact expressions for 
$N_{f,cr}$ for $S_2$ and $A_2$, we find 
\beq
N_{f,cr,SI}-N_{f,cr} = \frac{3N_c^3}{(N_c \pm 2)(5N_c^2 \pm 4N_c - 8)} \ .
\label{nfcrdif_t2}
\eeq
This difference is positive for all $N_c$ for both the $S_2$ and $A_2$ cases.
Thus, as with the fundamental and adjoint representations, for these rank-2
tensor representations, this perturbative approach overestimates $N_{f,cr}$ and
hence underestimates the size of the IR conformal phase.  A second perturbative
estimate of $N_{f,cr}$ is obtained by setting $\gamma_{IR,2\ell}=1$ and solving
for $N_f$ .  The condition $\gamma_{IR,2\ell}=1$ is a cubic equation in $N_f$,
from which we extract the physically relevant root.  This second method yields
estimates of $N_{f,cr}$ that are qualitatively similar to those obtained with
the first method with $\gamma_{IR,SI}$.  This qualitative agreement between
these two perturbative methods for these rank-2 tensor representations is the
same as what we found for the fundamental and adjoint representations.

% ======================================================================

\section{Discussion and Comparison with Non-supersymmetric SU($N_c$) Gauge
  Theory }

From our calculations on an SU($N_c$) gauge theory with ${\cal N}=1$
supersymmetry in this paper, we have found several general results.  Our most
detailed analyses here were for the cases $R=F$ and $R=Adj$, with briefer
studies of the $S_2$ and $A_2$ cases. It is useful to compare our results with
what we found in \cite{bvh} (see also \cite{ps}, whose results were in
agreement with those in \cite{bvh}) for a non-supersymmetric SU($N_c$) gauge
theory with $N_f$ copies of massless fermions in various representations.  We
believe that our findings for the supersymmetric gauge theory, besides being of
interest in their own right, provide further insight into the results that
we obtained previously for the non-supersymmetric theory.

First, for the IR zero of $\beta$, we find that $\alpha_{IR,3\ell} <
\alpha_{IR,2\ell}$.  This is the same type of shift that we showed earlier for
the non-supersymmetric theory with the same $R$ and suggests that the
lowest-order (two-loop) perturbative calculation of the IR fixed point gives a
larger value than the true value.  Second, we find that when one goes from the
two-loop anomalous dimension evaluated at the two-loop IR zero of $\beta$,
$\gamma_{IR,2\ell}$ or the scheme-independent $\gamma_{IR,SI}$, to the
three-loop result $\gamma_{IR,3\ell}$, the value decreases.  Again, this is the
same trend that we found for the corresponding non-supersymmetric theory in
\cite{bvh}.  Thus, as with the IR zero, this suggests that for both the
non-supersymmetric and the supersymmetric theory with corresponding matter
field representation content, the lowest-order calculation of the value of
$\gamma_m$ at the IR fixed point gives a larger value than the true value.  The
exact value of $N_{f,cr}$ is not known for the non-supersymmetric theory, and
an intensive research program has been underway for several years, especially
using lattice measurements, to determine $N_{f,cr}$ for a given $N_c$ and $R$.
Here we have taken advantage of the fact that $N_{f,cr}$ is known exactly (at
least with the level or rigor that is usual in physics) for the supersymmetric
SU($N_c$) theory.  We have used one method for obtaining a perturbative
estimate of $N_{f,cr}$ here, namely to set $\gamma_m=1$ and solve for the value
of $N_f$ where this occurs.  With this method, we have found that, for a given
$N_c$ and matter superfield content, the perturbative calculation yields a
slight overestimate of $N_{f,cr}$ as compared with the exactly known
value. This result agrees with and complements the different analysis in
Ref. \cite{ans98}, which was based on an analysis of an approximate solution to
the Dyson-Schwinger equation for the fermion matter field propagator. Our
calculation of $\gamma_m$ at the IR zero of the beta function provides some
insight into this.  Since, at least at the scheme-independent and two-loop
level, $\gamma_m$ increases throughout the IR conformal phase as $N_f$
decreases from $N_{f,max}$, and since the lowest-order calculations yield a
larger value of $\alpha_{IR}$ and $\gamma_m$ than the true value, it would
follow that setting $\gamma_m=1$ to determine the lower boundary of the IR
conformal phase would yield a value of $N_{f,cr}$ that is larger than the true
value.  One must, however, add the caveat that at the four-loop level for the
non-supersymmetric theory and at the three-loop level for the corresponding
supersymmetric theory, we have found that $\gamma_m$ does not increase
monotonically as $N_f$ decreases from $N_{f,max}$, which complicates the
intepretation of the results.  The deviation of $\gamma_{IR,4\ell}$ from
$\gamma_{IR,3\ell}$ in the non-supersymmetric theory was relatively small
throughout much of the $N_f$ interval of interest, but here we have found that
the deviation of $\gamma_{IR,3\ell}$ from $\gamma_{IR,2\ell}$ (or
$\gamma_{IR,SI}$) is significant in the $N_f$ region of interest, which limits
what one can infer from calculations of $\gamma_{IR,3\ell}$.

% ========================================================================

\section{Conclusions}

In this paper we have considered an asymptotically free vectorial SU($N_c$)
gauge theory with ${\cal N}=1$ supersymmetry and $N_f$ pairs of chiral
superfields $\Phi_i$, \ $\tilde \Phi_i$, $i=1,...,N_f$, transforming according
to the representations $R$ and $\bar R$, respectively, where $R$ includes
fundamental, adjoint, and symmetric and antisymmetric rank-2tensor
representations. We have studied the evolution of this theory from the
ultraviolet to the infrared, taking account of higher-loop corrections to the
$\beta$ function and the anomalous dimension $\gamma_m$.  We have compared the
results obtained from the two- and three-loop calculations of the beta function
and anomalous dimension (in the $\overline{DR}$ scheme) with exact results. In
particular, we have calculated perturbative estimates for the lower boundary of
the conformal phase, $N_{f,cr}$, by setting the scheme-independent and two-loop
perturbative expressions for $\gamma_m$ equal to the rigorous upper bound
(\ref{gammabound}), taken to be saturated at $N_{f,cr}$.  We find that this
perturbative calculation somewhat overestimates $N_{f,cr}$ as compared with the
exact results, and thus underestimates the size of the IR conformal phase.
Keeping in mind the caution that perturbative calculations become less reliable
as the infrared fixed point $\alpha_{IR}$ gets larger, our results provide a
measure of how closely perturbative calculations reproduce exact results
for these theories.

\begin{acknowledgments}

This research was partially supported by the {\it Sapere Aude} Award (TA) and 
the grant NSF-PHY-09-69739 (RS). 

\end{acknowledgments}


\begin{thebibliography}{99}
\bibitem{az}
%
We recall that the lack of analyticity at $\alpha=0$ is clear, for example,
from the fact that if the expressions were analytic at $\alpha=0$, then one
could analytically continue them from positive to negative $\alpha$, but this
would lead to an imaginary gauge coupling, which would render the theory 
non-unitary.
\bibitem{nfintegral}
%
Here and elsewhere, when expressions are given for $N_f$ that evaluate
to non-integral real values, it is understood that they are formal and are to
be interpreted via an analytic continuation of $N_f$ from physical integer
values to real values.
\end{thebibliography}
\end{document}